\documentclass[twocolumn,tightenlines,showpacs,prd,floatfix,preprintnumbers,amsmath,amssymb]{revtex4}

\usepackage{graphicx}

\usepackage{bm}

\begin{document}

\title{Cosmic String Cusps with Small-Scale Structure: Their Forms and
Gravitational Waveforms}

\author{Xavier Siemens}
\affiliation{Center for Gravitation and Cosmology, 
Department of Physics,University of Wisconsin --- Milwaukee,
P.O. Box 413,
Wisconsin, 53201, USA}

\author{Ken D. Olum}
\affiliation{Institute of Cosmology,
Department of Physics and Astronomy,
Tufts University, Medford, MA 02155, USA}

\date{\today}

\begin{abstract}
We present a method for the introduction of small-scale structure into
strings constructed from products of rotation matrices.  We use this
method to illustrate a range of possibilities for the shape of cusps
that depends on the properties of the small-scale structure.   We
further argue that the presence of structure at cusps under most
circumstances leads to the formation of loops at the size of the
smallest scales.  On the other hand we show that the gravitational
waveform of a cusp remains generally unchanged; the primary effect of
small-scale structure is to smooth out the sharp waveform emitted in
the direction of cusp motion.
\end{abstract}

\preprint{WISC-MILW-03-TH-1}

\maketitle

\section{Introduction}

The study and detection of gravitational waves is a very active area
of current research. Gravitational radiation is a direct test of
dynamical gravity and elucidating its properties experimentally has
the potential to confirm (or even change) our current gravitational
theory. Furthermore, the detection of a gravitational wave background
would open a new observational window onto a time in the early
universe much earlier than that of recombination, beyond which the
universe is opaque to electromagnetic waves.  The potential rewards to
cosmology that would result from such a detection are countless.  With
the new interferometric gravitational wave telescopes (LIGO, GEO and
TAMA) now operating and plans for a future space-based telescope
(LISA) it will become possible to detect waveforms from various
astrophysical sources, including topological defects. This has the
potential to determine, or at worst constrain, the types of high
energy particle theories that describe our world.

Topological defects are a prediction of most particle physics models
that involve a symmetry breaking phase transition
\cite{kibble76,alexbook} as well as some brane-world scenarios
\cite{SarangiTye} and are therefore quite generic.  Cosmic strings, in
particular, have drawn considerable attention because they do not
cause cosmological disasters and are good candidates for a variety of
interesting cosmological phenomena such as gamma ray bursts \cite{4},
gravitational wave bursts \cite{5} and ultra high energy cosmic rays
\cite{6}.

Analytic studies \cite{kibble85,bennet86} and numerical simulations
\cite{at,bb,as} show that cosmic string networks evolve toward an
attractor solution known as the ``scaling regime''. In this regime the
energy density of the string network is a small constant fraction of
the radiation or matter density and the statistical properties of the
system such as the correlation lengths of long strings and average
sizes of loops scale with the cosmic time \( t \). This solution is
possible because of intercommutations that produce cosmic string
loops which in turn decay by radiating gravitationally.
 
Simulations also have found that the perturbations on long strings and
most loops have the smallest possible size, the simulation resolution,
which does not scale. This small-scale structure arises from a
build-up of the sharp edges (kinks) \cite{AllenCald} produced at
string intersection events and the initial Brownian character of the
network. It is not clear, however, how very small loops can be formed
at late times. The mechanism that has so far been suggested is that
they form at regions of the string known as cusps \cite{albrecht},
although in this case we expect the loops formed to have relativistic
speeds. This is discussed at some length below. It may also be that
the small loops form from the collapse of horizon-sized loops. The
size of this small-scale structure is also thought to scale with the
cosmic time \( t \). The value typically quoted in the literature,
first proposed in \cite{bb2}, is \( \Gamma G\mu t\), where \( \Gamma
\) is a number of order 100, \( G \) is Newton's constant and $\mu$ is
the mass per unit length of the string, but the actual value
may be many orders of magnitude smaller \cite{us}.

One of the key signatures of cosmic strings is the production of
gravitational waves \cite{alex1}.  The network of long strings and loops is
expected to produce a gravitational wave background as well as bursts
coming from cusps and kinks. These bursts stand out of the
gravitational wave background and may be detectable with current
sensitivities \cite{5}.

In this paper we investigate the shape and gravitational waveforms of
cusps with small-scale structure.  In the next section we review the
motion of strings and construct solutions where helical small-scale
structure is introduced using products of rotations. In Section III we
investigate various small-scale structure regimes and show that the
presence of small-scale structure at cusps under most circumstances
leads to the production of loops at the size of the smallest
scales. We further illustrate the various regimes, and shapes of the
resulting cusps, using the solutions found in Section II. In Section
IV we describe the numerical algorithm employed to compute
gravitational waveforms and in Section V we show the waveforms for
cusps on loops with and without small-scale structure. We conclude in
Section VI.

\section{String motion}

When the typical length scale of a cosmic string is much larger than
its thickness, and long-range interactions between different string
segments can be neglected, the string can be accurately modeled by a
one-dimensional object. Such an object sweeps out a two-dimensional
surface in space-time referred to as the string world-sheet. Here we
will consider the motion of strings in flat space, which is a good
approximation to motion in a cosmological setting provided we are
studying a loop or a segment on a long string which is much smaller
than the horizon.

To study the motion of strings we use the Nambu-Goto action
\cite{nambu,goto} which is proportional to the area of the string
world-sheet. Minimising this action in the gauge where
\begin{equation}
\label{cont}
\partial _{u} x ^{\mu} \partial _{u} x _{\mu} =
\partial _{v} x ^{\mu} \partial _{v} x _{\mu} =0.
\end{equation}
yields the wave equation for the string
\begin{equation}
\label{cont2}
\partial _{u} \partial _{v} x ^{\mu}(u,v) =0.
\end{equation}
The coordinates $u$ and $v$ are null on the string world-sheet and
typically given in the literature as $u=t-\sigma$ and $v=t+\sigma$,
where $t$ is the time coordinate and $\sigma$ is a spatial parameter
that measures energy density along the string.  The constraint
Eqs.~(\ref{cont}), leave some residual freedom which can be removed by
imposing $a^0=u$ and $b^0=v$.

The general solution to Eq.~(\ref{cont2}) can be therefore be written
as
\begin{equation}
{\bf x} (\sigma ,t)= \frac{1}{2} [{\bf a} (t-\sigma)
+{\bf b} (t+\sigma)]
\end{equation}
and the constraint Eqs. (\ref{cont}) become 
\begin{equation}
{{\bf a}'}^2 =
{{\bf b}'}^2 = 1,
\label{constr}
\end{equation}
where ${{\bf a}'}$ and ${{\bf b}'}$ are the derivatives of ${{\bf a}}$
and ${{\bf b}}$ with respect to their arguments, $t-\sigma$ and
$t+\sigma$ respectively.  These vectors therefore live on a unit
sphere \cite{KT82} and will be referred to as the right- and
left-moving halves of the string.

Expressions for the functions ${{\bf a}'}$ and ${{\bf b}'}$ are most
easily written as sums of Fourier harmonics. Unfortunately the unit
magnitude constraint Eq.~(\ref{constr}) generally gives a non-linear
set of equations involving the vector coefficients of the Fourier
expansion and parametrising strings beyond the first few harmonics
proves to be analytically intractable. There exists a method to
generate strings involving products of rotation matrices that greatly
simplifies this task \cite{BrownDelaney}.  In this method the ${\bf
a}'$ and ${\bf b}'$ are generated by starting with a unit vector and
operating on it with a series of rotation matrices that through
trigonometric identities produce harmonics.  The advantage of using
this method to generate loops is that the unit magnitude constraint on
the right- and left-movers is satisfied trivially.

As an example of this method we can define the following loop
\begin{equation}
\label{simpleloop}
{\bf a}'(u)=R_{xy}(\alpha)R_{yz}({2\pi n \over L}u){\hat z}, \:
{\bf b}'(v)=R_{xz}({2\pi m\over L}v){\hat z},
\end{equation}
where $R_{ij}(\gamma)$ is a $3$-dimensional rotation matrix that
rotates the $i$-$j$ plane by an angle $\gamma$; the direction of the
rotation is chosen such that the positive end of the $i$th axis is
rotated toward the positive end of the $j$th axis.  On the unit
sphere, ${\bf a}'$ as given by Eq.~(\ref{simpleloop}) is a unit circle
which is tilted away from the $y$-$z$ plane by an angle
$\alpha$. Similarly ${\bf b}'$ is a unit circle on the $x$-$z$ plane.
This loop is (aside from an overall spatial rotation) identical to the family
of loops introduced by Burden \cite{Burden}.

Another advantage of the method of products of rotations is that it
lends itself readily to modifications of the loop trajectory.  It is
particularly easy to introduce helical perturbations in one or both
directions.  If we examine Eq.~(\ref{simpleloop}) we observe that both
${\bf a}'$ and ${\bf b}'$ consist of a rotation of the unit vector
${\hat z}$ on some plane.  To introduce helical perturbations in, say,
${\bf b}'$, we can replace the starting unit vector with
\begin{equation}
\label{wigglyunit}
{\hat z} \rightarrow {\hat z}'=
(1-\epsilon_b^2)^{1/2}{\hat z}+\epsilon_b R_{xy}
({2 \pi l_b \over L}v) {\hat x},
\end{equation}
with $\epsilon_b < 1$ and then subject this new unit vector to a
rotation on the same plane as before, namely,
\begin{eqnarray}
\label{wigglybp}
{\bf b}'(v)&=&R_{xz}({2\pi m\over L}v) {\hat z}'
\nonumber
\\
&=&R_{xz}({2\pi m\over L}v) 
((1-\epsilon_b ^2)^{1/2}{\hat z}+\epsilon_b R_{xy}({2 \pi l_b \over L}v) 
{\hat x}).
\nonumber
\\
\end{eqnarray}
The unit vector given by Eq.~(\ref{wigglyunit}) consists of two parts.
The first part is a vector of magnitude $(1-\epsilon_b ^2)^{1/2}$ in
the direction of ${\hat z}$ and it corresponds to the straight part of
${\bf b}'$. The second part is a circle of radius $\epsilon_b$ on the
$x$-$y$ plane, perpendicular therefore to ${\hat z}$, that winds
around this plane $l_b$ times. This represents the perturbed part of
${\bf b}'$.  Acting on this new unit vector ${\hat z}'$ with the
rotation $R_{xz}({2\pi m\over L}v)$ produces a curve in the unit
sphere with helical perturbations of radius $\epsilon_b$.  An
analogous operation can be carried out on ${\bf a}'$.

\begin{figure}
{\centering
\resizebox*{1.1\columnwidth}{0.35\textheight}
{\includegraphics{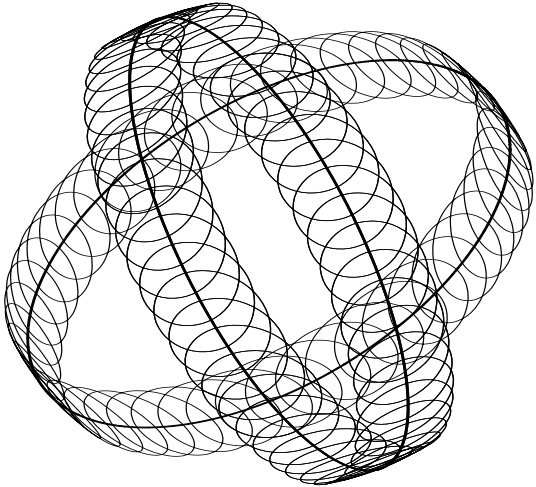}} \par}
\caption{
Right and left movers on the unit sphere. The thicker curves that
encircle the sphere once correspond to ${\bf a}'$ and ${\bf b}'$ as
given by Eq.~(\ref{simpleloop}) with $\alpha=0$, $n=1$ and $m=2$.  The
complex curves correspond to the perturbed versions of ${\bf a}'$ and
${\bf b}'$ with $\epsilon_a=\epsilon_b=0.2$, $l_a=100$ and $l_b=200$.
We have introduced twice as many oscillations in the perturbation for
${\bf b}'$ because the unperturbed ${\bf b}'$ winds twice around the
unit sphere.}
\label{fig:unit1}
\end{figure}

Figure \ref{fig:unit1} illustrates the effect of this
procedure. Equation (\ref{simpleloop}) gives ${\bf a}'$ and ${\bf b}'$
that trace out great circles on the unit sphere.  Perturbations from
Eq.~(\ref{wigglybp}) and its analog for ${\bf a}'$ add small, circular
oscillations to the larger loops.

\section{The forms of cosmic string cusps in the presence 
of small-scale structure}

Cusps are regions on long strings or loops that achieve phenomenal
Lorentz boosts. These boosts, combined with the potentially large mass
per unit length of cosmic strings, may lead to the production of a
detectable gravitational wave signal \cite{5}.

Cusps arise when the right- and left-moving parts of the string cross
on the unit sphere. At the crossing they point in the same direction,
namely,
\begin{equation}
{{\bf a}'} =
{{\bf b}'}.
\label{cusp1}
\end{equation}
In this case
\begin{equation}
{{\bf x}'} =\frac{1}{2}[-{{\bf a}'}+{{\bf b}'}]=0,
\label{cusp2}
\end{equation}
and
\begin{equation}
{\dot {\bf x}} =\frac{1}{2}[{{\bf a}'}+{{\bf b}'}]={{\bf a}'}={{\bf b}'}.
\label{cusp3}
\end{equation}
Here ${{\bf x}'} =\partial_{\sigma}{\bf x}$ and ${\dot {\bf x}}
=\partial_t{\bf x}$.  The tip of the cusp therefore moves at the speed
of light and a small region of string around the cusp has an enormous
Lorentz boost. Cusps are generic in the sense that the ${\bf a}'$ and
${\bf b}'$ are confined to live on the unit sphere and for general
string and loop configurations we typically expect them to cross.

The effects on cusps of the presence of small-scale structure on just
one of the string halves (`chiral' structure) was investigated briefly
in \cite{SandO}.  There it was argued that the shape of the cusp
depends on the contribution of the small perturbations to the
effective mass per unit length of the string. Here we will review and
expand that discussion to the more general case of small-scale
structure on both halves of the string. Although the detailed
properties of small-scale structure in cosmic string networks are
still not understood it is useful to study the range of possibilities.
Below we show that under most circumstances the presence of small-scale
structure at a cusp leads to the formation of loops at the size of
the smallest scales.

For simplicity we consider a loop of size $L$ with small-scale
structure of wavelength $\lambda$ and amplitude to wavelength ratio
$\epsilon$. For now we take the small-scale structure to be the same
in both directions.  The string has an effective thickness due to the
small-scale structure,
\begin{equation}
\label{deff} d\sim 2\epsilon \lambda.
\end{equation}

When a long string or loop has a cusp the effective thickness of the
string may result in overlap. Calculations for the overlap of the
fields making up the string core have been performed in
\cite{brandenberger} and \cite{KenandJose}.  In
the case of core overlap the fields in a small section of the string
near the tip of the cusp unwind and produce energy in the form of
particles, the bosons that make up the string. Fortunately we can
borrow these results, expecting, however, for overlap to result in the
formation of cosmic string loops rather than particles.  The latter
work \cite{KenandJose} took account of the Lorentz contraction of the
core and is the correct calculation for generic string core overlap.
The former work \cite{brandenberger} did not account for the
contraction of the string core but, as we will see, the main result
may be useful in the case of back-reaction-dominated cusps.

If the small-scale structure is small enough to have little effect on
the average motion of the string, then the analysis of
\cite{KenandJose} applies, and we expect an overlapping region near
the cusp of length
\begin{equation}
\label{loverlap}
l\sim \sqrt{d L}
\end{equation}
which will typically fragment into about $\sqrt{\epsilon L/\lambda }$ relativistic
loops of the small-scale size $\lambda$.  Here the effect of the
small-scale structure has been to replace the core thickness of the
string with the effective thickness $d$, which is usually much larger.

The overlap begins when the Lorentz gamma
factor of the cusp reaches the value
\begin{equation}
\label{gammaoverlap} \gamma_o \sim \sqrt{L/d}.
\end{equation}
When the small-scale structure is not symmetrically distributed
between the right- and left-movers at the cusp, the overlap
expressions, Eqs.~(\ref{loverlap}) and (\ref{gammaoverlap}), will be
dominated by the larger of the two halves.

On the other hand, if the small-scale structure is large enough, there
will be important corrections to the Nambu-Goto motion.  The effect of
small-scale structure increases as the string begins to move rapidly
near a cusp, with the effective mass per unit length becoming
\begin{equation}
\label{mueff1} \mu _{\rm eff}=\mu (1+\gamma ^{2}\epsilon ^{2}).
\end{equation}
Therefore when the gamma factor reaches the value
\begin{equation}
\label{gammabackreact} \gamma_b \sim 1/\epsilon
\end{equation}
the contribution to the effective mass per unit length of the string
becomes significant and the effective motion of the string deviates from
regular Nambu-Goto motion.  Averaging the motion of the string over
scales of the size of the small-scale structure and smaller results in
an increase of the effective mass density and a decrease of the
tension of the string \cite{12}.

Whether back-reaction prevents the formation of the relativistic cusp
depends on which one of the two gamma factors,
Eq. (\ref{gammabackreact}) or Eq. (\ref{gammaoverlap}), is the
smallest. If $\epsilon^2  L/d < 1$, then $\gamma_b >\gamma_0$, and
overlap begins before back-reaction is important.  On the other hand,
if $\epsilon^2  L/d > 1$, then back-reaction changes the form of the cusp.

The simplest situation of the latter sort occurs when both string halves
contribute equally to the effective mass per unit length.  Then the
effective motion of the string is given by \cite{alexbook}
\begin{equation}
\label{backreactmotion} 
{\bf x} (\sigma ,t)= \frac{1}{2} [{\bf a} (k u) +{\bf b} (k v)]
\end{equation}
with $k=\sqrt{1-\epsilon^2}$. The two halves therefore live on a
sphere of radius $k< 1$.  Sharp cusps form as before, because at
crossings on the sphere the effective ${\bf x}'=0$, but the tips of
cusps move at a speed $|{\dot {\bf x}}|=k$ rather than at the speed of
light.  Because the cusps are sharp we expect the small-scale
structure to overlap. If the perturbations are small then the cusp is
relativistic and we expect an overlapping region around the cusp of a
size given by Eq.~(\ref{loverlap}) and the production of relativistic
loops. However, if the small-scale structure is sufficiently large the
region of overlap will be larger due to the negligible Lorentz
contraction of the core and we can use the result in
\cite{brandenberger}. In this case we expect an overlapping region of
size
\begin{equation}
\label{loverlap2}
l\sim d^{1/3} \,L^{2/3}
\end{equation}
around the cusp, which will fragment into about $\epsilon^{1/3}
(L/\lambda)^{2/3}$ loops of the small-scale size $\lambda$.  Since the
loops produced in this case are non-relativistic, this mechanism may
be responsible for the large number of small loops seen in
simulations.

At first glance the small-scale structure in numerical simulations
would appear too small to slow down cusps sufficiently to produce
non-relativistic loops.  However, the wavelength of the perturbations
on long strings is a very small fraction of the length of the long
strings and the fact that it can be observed at all in simulations
means the amplitude to wavelength ratios cannot be negligibly small.

If the small-scale structure is not symmetrically distributed between
the right- and left-movers at the cusp, for example because of
statistical fluctuations, the radii of the spheres on which the
effective right- and left-moving halves live will be different. They
will therefore miss each other and make ${\bf x}'$ finite, with
\begin{equation}
|{\bf x}'|\sim |k_b-k_a |\sim |\epsilon_a^2-\epsilon_b^2 |
\end{equation}
at the cusp.  Here, the sub-indices $a$ and $b$ distinguish between
right- and left-movers.  The limiting situation of `chiral'
small-scale structure (i.e., $\epsilon_b = 0$) was investigated in
\cite{SandO} and is similar to that of a superconducting string with a
chiral neutral current \cite{carterpeter, jose}.

Since ${\bf x}'$ is finite the cusp will be rounded.  If the radius of
curvature of the cusp is comparable to the effective thickness of the
string then we are in the same situation as before: a section of
string will overlap near the cusp.  The amount of overlap will be
given by either one of Eqs.~(\ref{loverlap}) or (\ref{loverlap2})
depending on the size of the small-scale structure.

If the radius of curvature of the cusp is much larger than the
thickness then there will be no overlap. This case is analogous to
that of superconducting strings with a chiral neutral current and we
expect self-intersections to occur at the cusp about half of the time
\cite{13}. These self-intersections result in the production of a loop
of size
\begin{equation}
l \sim |\epsilon_a^2-\epsilon_b^2|^{1/2} L
\end{equation}
which will fragment into about
$|\epsilon_a^2-\epsilon_b^2|^{1/2}L/\lambda$ loops of the small-scale
size $\lambda$.

To summarise, we have seen that in most cases overlap of the
small-scale structure at a cusp leads to the formation of loops at the
smallest scales. The only exception occurs when the small-scale
structure is a) sufficiently large that the averaged motion of the
string deviates significantly from Nambu-Goto motion, and b)
sufficiently asymmetric that the radius of curvature of the cusp is
larger than the effective thickness of the string. Even in this case,
however, we expect intersections to produce small loops about half the
time.

\begin{figure}
{\centering
\resizebox*{1.\columnwidth}{0.3\textheight}
{\includegraphics{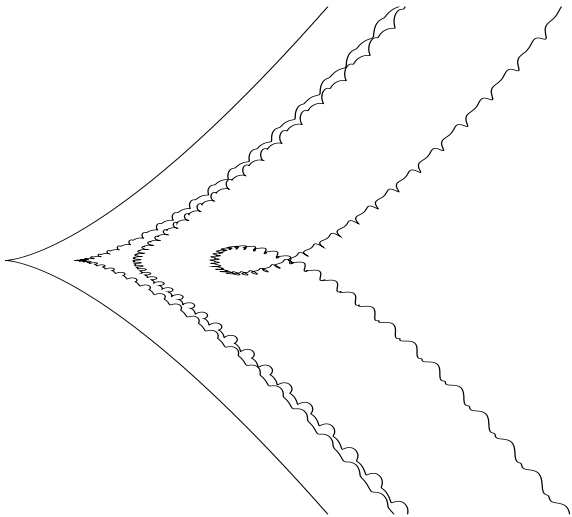}}
\par} \caption{
On the left, a projection of an unperturbed cusp on a loop given by
Eq.\ (\ref{simpleloop}) with $\alpha=0$, $n=1$ and $m=2$.  Next to
the right, a perturbed version of the same cusp with
$\epsilon_a=\epsilon_b=0.3$, $l_a=200$ and $l_b=400$.
Next a projection of a cusp with $\epsilon_a=0.4$,
$\epsilon_b=0.5$, $l_a=200$ and $l_b=400$. In this case the effective
${\bf x}'$ is finite and the cusp is smoothed out. On the right,
a cusp with $n=2$, $m=1$, $\epsilon_a=0.4$, $\epsilon_b=0.5$,
$l_a=200$ and $l_b=400$.  Switching the values of $n$ and $m$ has
inverted the sign of ${\bf x}'$ and led to a self-intersection near
the cusp.} \label{fig:loop1close}
\end{figure}

Figure \ref{fig:loop1close} illustrates the various possibilities for
the shape of cusps with small-scale structure described above.
Without small-scale structure, we have a sharp cusp which would
evaporate at the tip by core overlap.  With symmetrical small
excitations, the cusp is still sharp and will chop off a loop by the
overlap of the wiggles.  With asymmetrical small-scale structure, the
cusp will be smoothed out.  There may or may not be an
intercommutation, depending on which half of the string has the larger
perturbations, as discussed in \cite{13}.

\section{Gravitational waveforms: formalism and numerical implementation}

Periodic sources produce a metric perturbation
that in the wave-zone is given by a sum of plane waves \cite{8}
\begin{equation}
\label{hmunu} h_{\mu\nu} ({\bf x},t)=\sum_{n=-\infty} ^{\infty}
\epsilon_{\mu\nu} ^{(n)} ({\bf x},\omega_n) e^{-ik_n \cdot x}.
\end{equation}
The polarisation tensor $\epsilon_{\mu\nu} ^{(n)}$ is given by
\begin{equation}
\label{polariz}
 \epsilon_{\mu\nu} ^{(n)} ({\bf x},\omega_n)={4G \over r} (
T_{\mu \nu}(k_n)-{1\over 2} \eta_{\mu \nu} T^\lambda _\lambda(k_n) )
\end{equation}
and 
\begin{equation}
\label{TmunuFTagain1} T^{\mu \nu }(k_n)=\frac{1}{T} \int_0 ^{T} dt
\int d^{3}{\bf x} T^{\mu \nu }(x)e^{ik_n \cdot x}
\end{equation}
is the Fourier transform of the stress-energy tensor of the source.
In these expressions $r$ is the distance to the source,  
$T$ is the period of motion,
\begin{equation}
k^\mu _n=\omega_n(1,{\hat \Omega})
\end{equation}
is the wave-vector of the gravitational wave, ${\hat \Omega}$ is a
unit vector pointing from the source to the point of observation and
$\omega_n=2\pi n/T$ is the frequency.

In the gauge defined by Eq.~(\ref{constr}) the stress energy tensor of
the string can be put in the form
\begin{equation}
\label{TmunuLightConeGauge} T^{\mu \nu }(x)=\mu \int
dudv(x_{,u}^{\mu }x_{,v}^{\nu } +x_{,u}^{\nu }x_{,v}^{\mu })\delta
^{(4)}(x-x(\zeta )).
\end{equation}

We consider a loop of length $L$ which oscillates periodically with period
$T=L/2$. The Fourier transform of the stress-energy tensor of the loop is 
\begin{eqnarray}
\label{TmunuLightConeGaugeFTagain} 
T^{\mu \nu }(k_n)&=&\frac{2\mu}{L} \int_0 ^{L/2} dt \int d^{3}{\bf x} 
e^{ik_n \cdot x}\int dudv
\nonumber
\\
&\times&[x_{,u}^{\mu }x_{,v}^{\nu
}+x_{,u}^{\nu }x_{,v}^{\mu }] \delta ^{(4)}(x-x(u,v)).
\end{eqnarray}

The expression for the string position in terms of right- and
left-moving parts when substituted in
Eq.~(\ref{TmunuLightConeGaugeFTagain}) yields the symmetrised product
\begin{equation}
\label{TmunuSimplified2} T^{\mu \nu }(k_n)=\frac{\mu L}{2} (A^{\mu
}(k_n)B^{\nu }(k_n)+A^{\nu}(k_n)B^{\mu }(k_n)),
\end{equation}
where
\begin{equation}
\label{Aofk2} 
A^{\mu }(k_n)={1 \over L} \int ^{L }_{0}du
a'^{\mu }(u)e^{ik_n\cdot a(u)/2}, 
\end{equation}
and
\begin{equation}
\label{Bofk2} 
B^{\mu }(k_n)={1 \over L} \int ^{L
}_{0 } dv b'^{\mu }(v)e^{ik_n\cdot b(v)/2}.
\end{equation}
This means that we can write the polarisation tensor Eq.~(\ref{polariz}) as
\begin{eqnarray}
\label{polariz2}
\epsilon_{\mu\nu} ^{(n)} ({\bf x},\omega_n)&=&{2G \mu L\over r} 
[A_{\mu}(k_n)B_{\nu }(k_n)
+A_{\nu}(k_n)B_{\mu }(k_n)
\nonumber
\\
&-&\eta_{\mu\nu}A_\lambda (k_n) B^\lambda (k_n)].
\end{eqnarray}

We are free to choose coordinates such that the wave-vector lies on
the $z$-axis. In this case the only physical components of the
polarisation tensor can be shown to be $\epsilon^{(n)} _{11}\equiv
\epsilon^{(n)} _{+}$ and $\epsilon^{(n)} _{12} \equiv \epsilon^{(n)}
_{\times}$ \cite{8}, which are referred to as the $+$- and
$\times$-polarisations of the gravitational wave \cite{MTW}.

It is easy to show using integration by parts and the periodicity of
$a^{\mu }$ and $b^{\mu }$ that
\begin{equation}
\label{AandBofkperp} 
k_n ^{\mu}A_{\mu }(k_n)
=k_n ^{\mu} B_{\mu }(k_n)=0.
\end{equation}
Using this identity and Eq.~(\ref{polariz2}) one finds that the 
amplitudes of the
$+$- and $\times$-polarised waves are given by
\begin{equation}
\label{plus}
\epsilon^{(n)} _{+} = {2G \mu L\over r}
(A_x(k_n)B_x(k_n)-A_y(k_n)B_y(k_n))
\end{equation}
and
\begin{equation}
\label{cross}
\epsilon^{(n)} _{\times} = {2G \mu L\over r}
(A_x(k_n)B_y(k_n)+A_y(k_n)B_x(k_n))
\end{equation}
respectively.

The metric perturbation can therefore be written as
\begin{equation}
\label{hmunuplucross} 
h_{+/ \times} ({\bf x},t)=\sum_{n=-\infty} ^{\infty}
\epsilon_{+/ \times} ^{(n)} ({\bf x},\omega_n) e^{-ik_n \cdot x}.
\end{equation}
The n=0 mode in this sum corresponds to the static field and should
not be included in the sum if we are interested in the radiative part
of the metric perturbation.

In order to study the gravitational waveforms produced by arbitrary
loops we must resort to numerical methods. If we consider a loop
composed of some continuous functions ${\bf a}(u)$ and ${\bf b}(v)$ we
can always approximate it by choosing some discretisation scheme that
is sufficiently fine so that all the features of the loop are
resolved.  We can use this discretisation to compute the integrals
Eqs.~(\ref{Aofk2}) and (\ref{Bofk2}) and find the gravitational
waveform it produces.

Here we use the piece-wise linear discretisation \cite{CasperAllen}.
Allen and Ottewill \cite{AllenOtt} computed the gravitational
waveforms for a number of different loop trajectories and showed that
the in the piece-wise linear approximation the waveform for those
loops is in good agreement with their analytic calculations.
Furthermore, as they found, an advantage of using this particular
approximation in the calculation of gravitational waveforms is that
the integrals Eqs.~(\ref{Aofk2}) and (\ref{Bofk2}) can be performed
exactly. This result is computed explicitly below.

In the piece-wise linear discretisation both functions ${\bf a}(u)$
and ${\bf b}(v)$ are constructed out of $N_a$ and $N_b$ straight
segments joined at kinks. The values of the internal parameters $u$
and $v$ are continuous and are labeled by $u_j$ with
$j=0,1,...,N_a-2,N_a-1$, and a similar expression for $v_j$, at the
kinks.  Here we will take $N_a=N_b=N$ and space the internal
parameters evenly between the kinks ($u_{j+1}-u_j=v_{k+1}-v_k=\delta$
for all $j$ and $k$) for simplicity.  The functions ${\bf a}(u)$ and
${\bf b}(v)$ are defined at the kinks
\begin{equation}
\label{aibi} 
{\bf a}_j={\bf a}(u_j),\: {\bf b}_k={\bf b}(v_k)
\end{equation}
and the derivatives of these functions in the intervening straight
segments between the kinks
\begin{equation}
\label{apibpi} 
{\bf a}' _j={{\bf a}_{j+1}-{\bf a}_{j} \over \delta},
\: {\bf b}' _k={{\bf b}_{k+1}-{\bf b}_{k} \over \delta}.
\end{equation}

Here we will focus only on $A^{\mu }(k_n)$ because the situation for
$B^{\mu }(k_n)$ is completely analogous.  For piece-wise linear loops
the integral of $A^{\mu }(k_n)$ in Eq.~(\ref{Aofk2}) becomes a sum of
integrals between kinks
\begin{equation}
\label{Anofk} 
A^{\mu }(k_n)={1 \over L} \sum^{N-1} _{j=0} a'^{\mu } _j 
\int ^{u_{j+1}}_{u_{j}}du
e^{ik_n\cdot a(u)/2}.
\end{equation}
Since the string segments are are taken to be straight between the
kinks
\begin{equation}
\label{amuxi} 
a^{\mu }(u)=a^{\mu } _{j}+(u - u _j)a'^{\mu } _{j},
\end{equation}
the integrand in Eq.~(\ref{Anofk}) is a simple exponential. 
From here it is straightforward to show that
\begin{eqnarray}
\label{Anofkinteg} 
A^{\mu }(k_n)&=&\sum^{N-1} _{j=0} {2 a'^{\mu } _{j} \over i L \omega_n
(1-{\hat \Omega} \cdot {\bf a}'_{j})}
\nonumber
\\
&\times&[e^{i \omega_n (u_{j+1}-{\hat \Omega} \cdot {\bf a}_{j+1})/2}-
e^{i \omega_n (u_j-{\hat \Omega} \cdot {\bf a}_{j})/2}]
\end{eqnarray}
with a similar expression for $B^{\mu }(k_n)$.

\section{Gravitational waveforms from cusps}

Gravitational bursts produced by cusps have previously been studied
analytically by Damour and Vilenkin \cite{5}.  They found the metric
perturbation produced by a cusp to be
\begin{equation}
\label{approxh4}
h(t) \propto |t-r|^{1/3} 
\end{equation}
where $r$ is the distance to the source. 

\begin{figure}
{\centering \resizebox*{1.0\columnwidth}{0.29\textheight}
{\includegraphics[angle=-90]{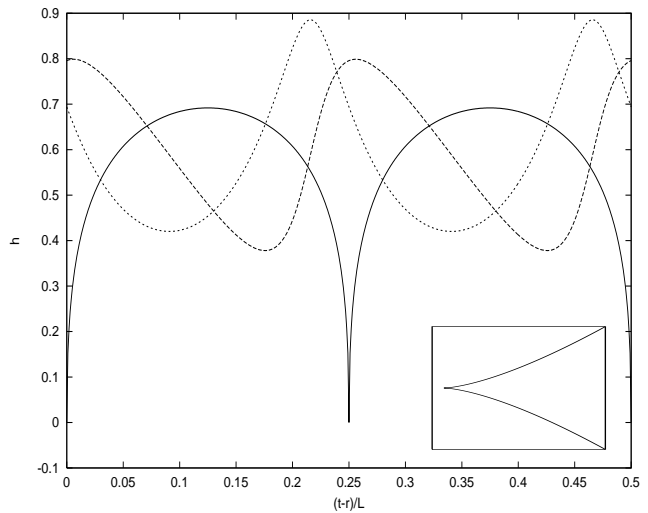}} \par} 
\caption{Waveforms for the loop defined by Eq.~(\ref{simpleloop})
with $\alpha=0$, $n=1$ and $m=2$, whose cusp is shown in the inset and
first in Figure \ref{fig:loop1close}, as a function of time.
The solid line is the $\times$-polarised waveform
when looking straight at the cusp in the $z$-direction.  No gravity
wave is emitted with $+$ polarisation in this direction.  For the same
loop with an overall spatial rotation of $1$ radian about the
$x$-axis, so that we are looking off-cusp, the $+$-polarised waveform
is dashed and the $\times$-polarised waveform is dotted.  All
waveforms are summed over the first $2000$ Fourier modes with
$2G\mu L/r=1$.  The loop is approximated with $5000$ linear segments
giving a resolution of the internal parameters $\delta=2 \times
10^{-4}L$.}
\label{fig:waveformcusp}
\end{figure}

As an example we can check waveform of a cusp on a loop.  We take the
loop defined by Eq.~(\ref{simpleloop}) which has a cusp in the
$z$-direction at $t=0$ and $t=L/4$.  The on- and off-cusp
waveforms for this loop are shown
in Figure \ref{fig:waveformcusp} for $\alpha=0$, $n=1$ and $m=2$.

\begin{figure}
{\centering \resizebox*{1.0\columnwidth}{0.29\textheight}
{\includegraphics[angle=-90]{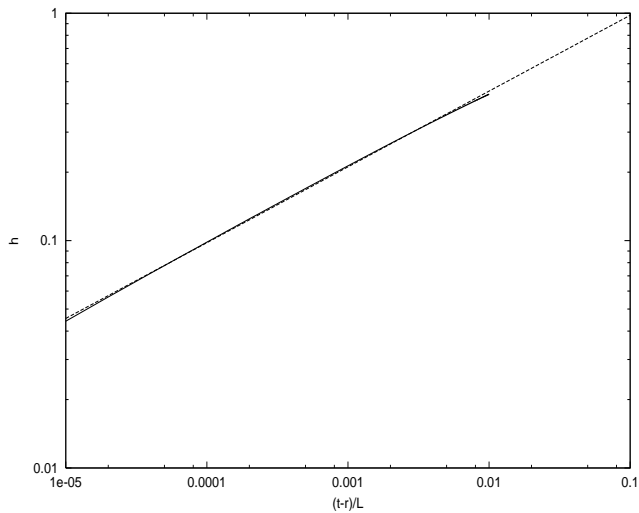}} \par} 
\caption{ 
Log-log plot of the cusp waveform for the loop defined by
Eq.~(\ref{simpleloop}) with $\alpha=0$, $n=1$ and $m=2$ as a function
of time. The solid curve gives the $\times$-polarised waveform
computed numerically and the dashed curve is given by
Eq.~(\ref{approxh5}) with $a=2.11$ (fit to the data).  The waveform
has been summed over the first $2\times 10^7$ Fourier modes with
$2G\mu L/r=1$.}
\label{fig:t13}
\end{figure}

Figure \ref{fig:t13} shows a log-log plot of the waveform versus time
near the cusp.  The solid curve gives the waveform computed numerically
over the first $2\times 10^7$ Fourier modes, with
$2G\mu L/r=1$ and the dashed curve is given by
\begin{equation}
\label{approxh5}
h_{\times}=a\,|t-r|^{1/3}
\end{equation}
with $a=2.11$ (fit to the data). The large number of Fourier modes
necessary to reproduce the behaviour of Eq.~(\ref{approxh4}) is due to
the slow decrease in amplitude of $h_{\times}$ as a function of the
Fourier mode, $h_{\times}(n)\propto n^{-4/3}$.

Due to the pervasiveness of small-scale structure in numerical
simulations of cosmic strings, it is important to test the robustness
of the prediction for the gravitational waveform of a cusp,
Eq.~(\ref{approxh4}), to the presence of small-scale structure.

\begin{figure}
{\centering
\resizebox*{1.0\columnwidth}{0.29\textheight}
{\includegraphics[angle=-90]{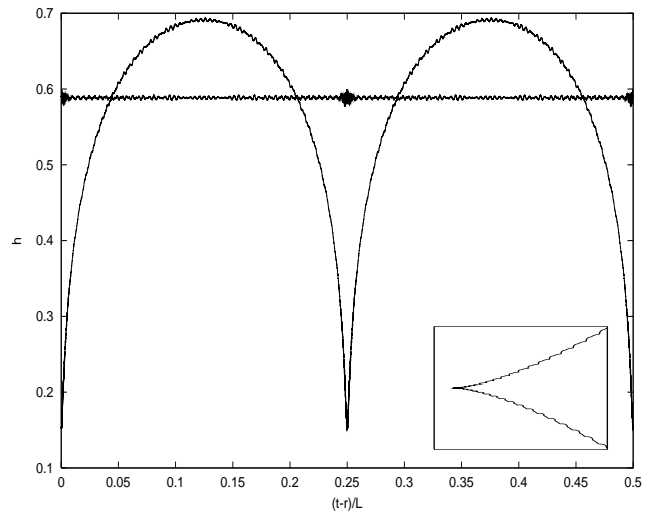}}
\par} \caption{
Waveforms for a perturbed loop, given by $\alpha=0$, $n=1$,
$m=2$, $\epsilon_a=\epsilon_b=0.3$, $l_a=200$ and $l_b=400$,
whose cusp is shown second in Figure
\ref{fig:loop1close} and in the inset, looking straight at the cusp
in the $z$-direction.  Because of the perturbations there is now a small
$+$-polarised wave and the form of the $\times$-polarised wave is not
as sharp as in Figure \ref{fig:waveformcusp}.} \label{fig:cuspwave2}
\end{figure}

\begin{figure}
{\centering
\resizebox*{1.0\columnwidth}{0.29\textheight}
{\includegraphics[angle=-90]{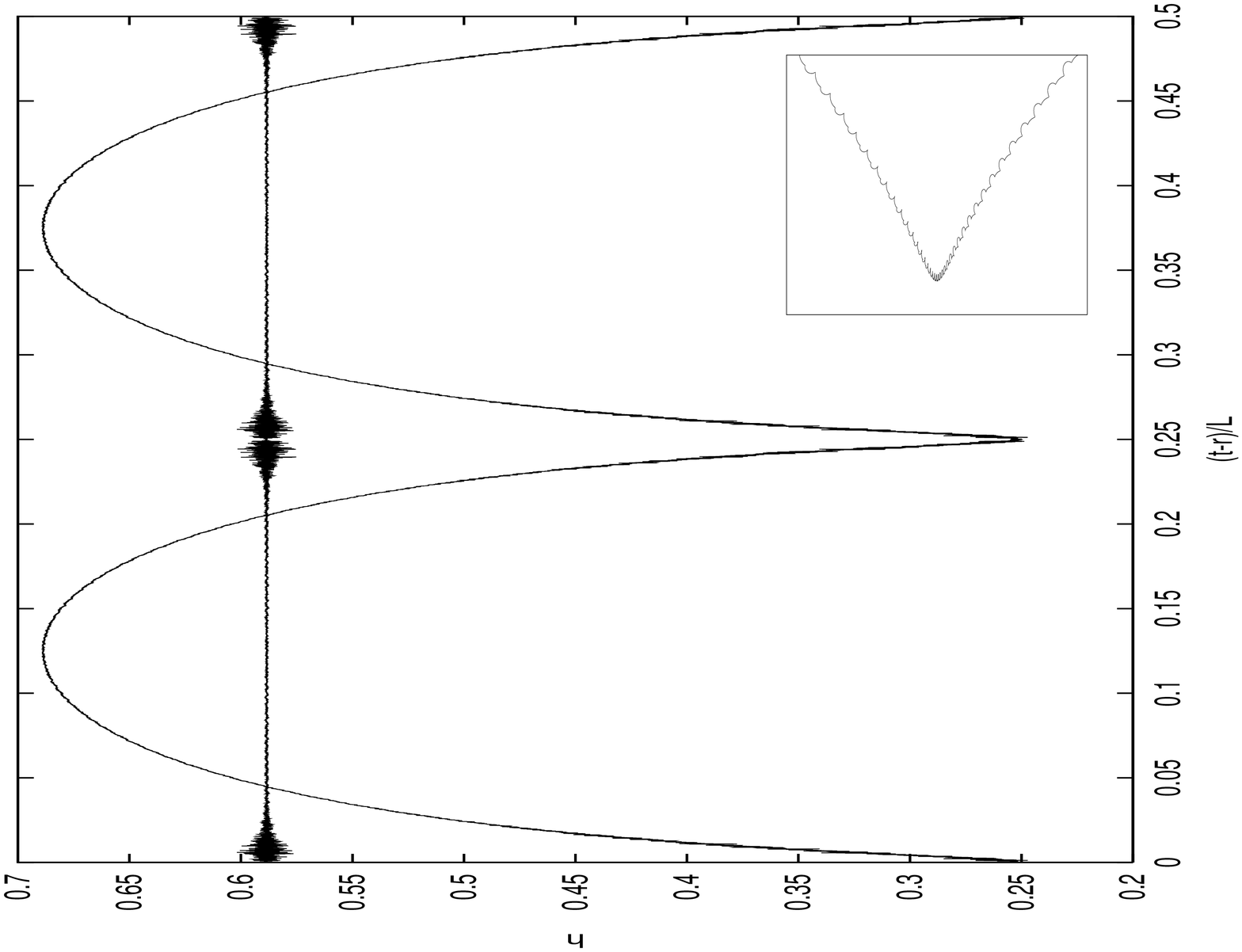}}
\par} \caption{
Waveforms for a perturbed loop with $\alpha=0$, $n=1$,
$m=2$,  $\epsilon_a=0.4$, $\epsilon_b=0.5$, $l_a=200$ and
$l_b=400$, whose cusp is shown third in Figure
\ref{fig:loop1close} and in the inset. 
} \label{fig:cuspwave3}
\end{figure}

\begin{figure}
{\centering
\resizebox*{1.\columnwidth}{0.29\textheight}
{\includegraphics[angle=-90]{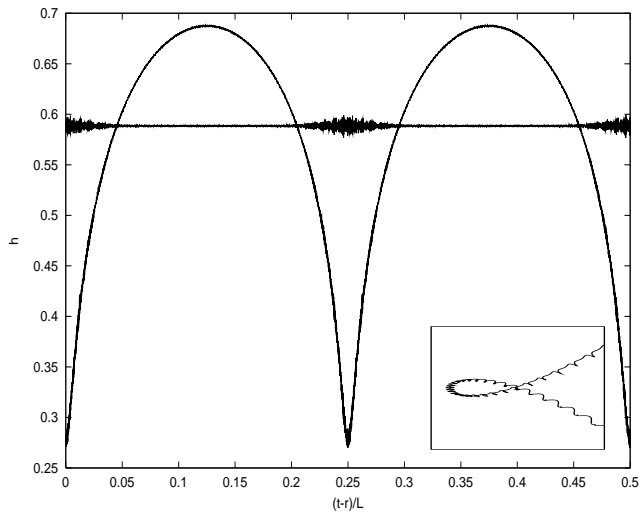}}
\par} 
\caption{
Waveforms for a perturbed loop with $\alpha=0$, $n=2$,
$m=1$, $\epsilon_a=0.4$, $\epsilon_b=0.5$, $l_a=200$ and
$l_b=400$, whose cusp is shown last in Figure
\ref{fig:loop1close} and in the inset.
}
\label{fig:cuspwave4}
\end{figure}

In order to investigate the effects of small-scale structure on the
gravitational waveforms of cusps we have performed a number of
simulations of cosmic string loops with small perturbations. 
Figures \ref{fig:cuspwave2}--\ref{fig:cuspwave4} show the waveforms of
the loops whose cusps appear in Figure~\ref{fig:loop1close}. The
waveforms are computed by summing over the first $10000$ Fourier
modes with $2G \mu L/r=1$.  The loops are approximated with
$5000$ linear segments.  Intercommutations are not performed in the
simulation; the strings just pass through each other without
producing new loops.

These figures show that the overall shape of the waveform of the cusp
is not greatly affected by the presence of small-scale structure even
when the amplitude of the small-scale structure $\epsilon$ is not
particularly small.  The primary effect is to smooth out the very
sharp point in the waveform.  But that point is visible only if the
observer is located in the direction of cusp motion.  Gravity waves
emitted from the string moving with Lorentz boost $\gamma$ are
confined to a cone with angle $\theta\sim\gamma^{-1}$
\cite{5}.  The smoothing of the sharp waveform is due
primarily to the limitation of the string boost at $\gamma_b$, so the
change in the waveform will be observed only if one is within angle
$\gamma_b^{-1}\sim\epsilon$ of the direction of cusp motion.

In addition to the smoothing of the large cusp the waveforms also show
significant sub-structure.  It is a result of the small helical
perturbations and is present in both the the $+$-polarised and
$\times$-polarised waveforms.

The sub-structure is particularly prominent at the time the large cusp
forms and corresponds to the waveforms of many tiny cusps. These
smaller cusps form as a result of the crossings on the unit sphere
that occur around the time the large cusp forms (as in Figure
\ref{fig:unit1}).  In terms of gravitational wave detection it is
possible that the small scale structure gives the dominant
contribution to the waveform in some range of frequencies. This
possibility depends on the value of $G\mu$ and the details of cosmic
string network evolution and deserves further investigation.

\section{Conclusions}

The presence of small-scale structure at cusps leads to a multitude of
interesting behaviours mostly involving self-intersections of the loop
or long string that produces the cusp. The specific processes and
outcomes depend on the detailed properties of the small-scale
structure. Indeed, one of the scenarios presented here supports the
ideas of Albrecht \cite{albrecht} who first suggested that the small
loops seen in simulations are produced at cusps.

At first glance it would appear that cusps are only able to produce
ultra-relativistic loops, which would rule out this mechanism since
such loops are not seen in simulations.  However, in the regime where the
small-scale structure has enough amplitude to significantly affect
the motion of the string, the cusp and the loops it produces
through overlap will no longer be relativistic.

As an illustration we take the canonical value for the size of the
small-scale structure, which for GUT strings is $\lambda \sim
10^{-4}t$, and a significant amplitude to wave-length ratio, $\epsilon
\sim 1$.  Then we expect a horizon-sized cusp, of size $L\sim t$, to
produce a few hundred non-relativistic loops
\begin{equation}
\label{loverlap3}
\epsilon^{1/3}(L/\lambda )^{2/3} \sim 500.
\end{equation}

The primary effect of small-scale structure is to smooth out the sharp
waveform emitted by the large cusp in the direction of its motion.
That waveform is smoothed anyway if one is located away from the exact
direction of motion; observers outside a cone whose angle is
proportional to the small-scale structure amplitude will see little change.
However, since superimposed on the large cusp there is a multitude of
smaller ones, it may be that the small-scale structure turns out to
give the dominant contribution to the gravitational wave signal in the
range of frequencies onto which we happen to have our observational
window.  This depends on the typical size of loops and the spectrum of
perturbations that ends up on loops. This possibility deserves further
investigation.

Unfortunately, the uncertainties that remain regarding string network
evolution make it difficult to say anything definite in this regard.
For instance, if it turns out to be correct that most loops are
produced at the smallest possible scales then we do not expect them to
have any significant structure.

\begin{acknowledgments}  
We would like to thank Alex Vilenkin for very helpful discussions and
suggestions and Bruce Allen and A. Ottewill for useful and friendly
correspondence. K. D. O. was supported in part by the National Science
Foundation. X. S. was supported by National Science Foundation 
grants PHY 0071028 and PHY 0079683.
\end{acknowledgments}

\end{document}